\documentclass{kapproc}
\usepackage{procps}
\usepackage[dvips]{graphicx}
 \normallatexbib

\begin{document}

\articletitle{Phase Transitions in Mesoscopic \hfill  \\
Superconducting Films}

\author{V. V. Kabanov}
\affil{J. Stefan Institute, Jamova 39,
1000 Ljubljana, Slovenia}
\email{viktor.kabanov@ijs.si}

\author{T. Mertelj}
\affil{Faculty of Mathematics and Physics,
Jadranska 19,
1000 Ljubljana, Slovenia \\and\\
J. Stefan Institute, Jamova 39,
1000 Ljubljana, Slovenia}

\email{tomaz.mertelj@ijs.si}

\begin{abstract}
We solve the Ginzburg-Landau equation (GLE) for the mesoscopic
superconducting thin film of the square shape in the magnetic field for the
wide range of the Ginzburg-Landau parameter $0.05<\kappa _{eff}<\infty $. We
found that the phase with the antivortex exists in the broad range of
parameters. When the coherence length decreases the topological phase
transition to the phase with the same total vorticity and a reduced symmetry
takes place. The giant vortex with the vorticity $m=3$ is found to be
unstable for any field, $\xi /a$ and $\kappa _{eff}\ge 0.1$. Reduction of $ \kappa _{eff}$ does not make the phase with antivortex more stable contrary
to the case of the cylindric sample of the type I superconductor.
\end{abstract}

\section*{Introduction}

\noindent Advances in nanotechnology and constantly shrinking
semiconductor devices have motivated researches to study
properties of mesoscopic superconducting samples. One line of
research in this field has focused on the problem of the phase
transitions in the mesoscopic superconducting sample under the
influence of the external magnetic field. \cite{mosch} We will
focus on the case when the size of the sample $a\sim \xi, \lambda
$, with $\xi $ and $\lambda$ being the superconducting coherence
length and the London penetration depth respectively. In that case
there are only a few vortices in the sample. The standard
Abrikosov approach \cite{abr} must be modified because of the
strong influence of sample boundaries. Thermodynamics of the
system is determined by the short-range repulsion of vortices and
interaction of vortices with boundaries.

Recently it was shown that the influence of boundaries can lead to
stabilization of the vortex-antivortex molecules in mesoscopic
samples.\cite {mosch} Analysis of the linearized Ginzburg-Landau
equation (GLE) has shown that such molecules appear at particular
values of the external magnetic field depending on the sample
shape and size\cite{chib}. The solution of the GLE in the limit of
the extreme type-II superconductor shows that such molecules have
a very shallow minimum in the free energy\cite{bonca,baelus} and
are very sensitive to the change of the sample shape\cite{meln}.

In a square mesoscopic thin film with the total vorticity $m=3$ the
symmetric solution with four vortices and one antivortex is the solution of
the linearized GLE with the lowest free energy\cite{mosch}. According to
ref. \cite{bonca}, away from the $H_{c2}$ line the giant vortex with
vorticity $m=3$ is stable and has the lowest free energy. This implies that
a topological phase transition \emph{without change} of the vorticity and
\emph{without a reduction} of the symmetry should take place with the change
of the external field or/and the coherence length away from the
critical-field line.

Here we report the results of the extensive studies of different
kinds of phase transitions for the thin film of the square shape.
We focused to the region $4<a/\xi <8$ and $\kappa _{eff}>0.05$
(see also\cite{mertelj}). Here $\kappa _{eff}=\lambda^2/d \xi$
where $d$ is the film thickness. We found that the antivortex
phase with $m=3$ is stable in a broad range of parameters. The
region of stability of the phase does not depend strongly on the
value of the parameter $\kappa _{eff}$. The energy gain due to the
antivortex formation is much smaller then the energy difference
between two phases with different vorticities. The giant vortex
with $m=3$ is unstable for any field, $\xi /a$ and $\kappa
_{eff}\ge 0.1$. Phase transition to the phase with three separated
vortices takes place when $\xi /a$ is driven away from the
critical field line. The reduction of $\kappa _{eff}$ does not
stabilize the antivortex phase for the thin film sample in the
contrast to the case of the cylindric sample ref.\cite{misko}.

\section{Formalism and Solution}

\noindent GLE for the normalized complex order parameter $\psi$
has the following form:\cite{bonca}
\begin{equation}
\xi ^{2}(i\nabla +{\frac{2\pi \mathbf{A}}{{\Phi _{0}}}})^{2}\psi -\psi +\psi
|\psi |^{2}=0
\end{equation}
here $\Phi _{0}$ is the flux quantum, $\mathbf{A}$ is the vector
potential and $\mathbf{H}=\nabla \times \mathbf{A}$ the magnetic
field. We split the vector potential into the external part due to
external currents, $\mathbf{A} _{ext}$, and the internal part due
to the response of the superconducting film, $\mathbf{A}_{int}$.
The second GLE equation for the total vector potential reads:
\begin{equation}
\nabla \times \nabla \times \mathbf{A}=-i{\frac{\Phi _{0}}{{4\pi \lambda ^{2}%
}}}(\psi ^{*}\nabla \psi -\psi \nabla \psi ^{*})-{\frac{|\psi |^{2}\mathbf{A}%
}{{\ \lambda ^{2}}}}.
\end{equation}
In addition to Eq.(1) we assume the boundary condition for the
superconductor insulator junction on the sample edges:
\begin{equation}
(i\nabla +{\frac{2\pi \mathbf{A}}{{\Phi _{0}}}})\cdot \mathbf{n}\psi =0,
\end{equation}
where $\mathbf{n}$ is the vector normal to the surface of the sample.

As it was described in ref.\cite{bonca} we introduce $N\times N$
discrete points on the square and rewrite Eq. (1) in the form of
the nonlinear discrete Schr\"{o}dinger equation:
\begin{equation}
\sum_{\mathbf{l}}t_{\mathbf{i+l,i}}\psi _{\mathbf{i+l}}-\epsilon (\mathbf{i}%
)t_{\mathbf{i,i}}\psi _{\mathbf{i}}-\psi _{\mathbf{i}}+\psi _{\mathbf{i}%
}|\psi _{\mathbf{i}}|^{2}=0,
\end{equation}
where the summation index $\mathbf{l}=(\pm 1,0)$, $(0,\pm 1)$ points toward
the nearest neighbors and $t_{\mathbf{i_{1},i}}=(\xi N/a)^{2}\exp(i\phi _{%
\mathbf{i_{1},i}})$ and $\phi _{\mathbf{i_{1},i}}=-{\frac{2\pi }{{\Phi _{0}}}%
}\int_{\mathbf{r}_{\mathbf{i}}}^{\mathbf{r}_{\mathbf{i_{1}}}}\mathbf{A}(%
\mathbf{r})d\mathbf{r}$. The boundary conditions are included in the
discrete nonlinear Schr\"{o}dinger equation as in ref.\cite{bonca} where $%
\psi _{\mathbf{i}}=0$ if $\mathbf{i}$ is outside of the sample and $\epsilon
(\mathbf{i})=4-\delta _{i_{x},1}-\delta _{i_{x},N}-\delta _{i_{y},1}-\delta
_{i_{y},N}$ where $\mathbf{i}=(i_{x}=1,\dots ,N,i_{y}=1,\dots ,N)$.

After discretization of Eq. (2) we obtain the exact expression for
the internal part of the vector potential\cite{mertelj}:
\begin{equation}
A_{int,\mathbf{i}}^{v}=\sum_{\mathbf{n}}K(\mathbf{i-n})J_{\mathbf{n}}^{v},
\end{equation}
where
\begin{eqnarray}
J_{\mathbf{i}}^{v} &=&\frac{\Phi _{0}a}{4\pi \lambda _{eff}N}\Im
(\exp(-i\phi _{\mathbf{i+l_{v},i}})\psi _{\mathbf{i}}^{*}\psi
_{\mathbf{i+l_{v}}}-
\nonumber \\
&&\exp(-i\phi _{\mathbf{i-l_{v},i}})\psi _{\mathbf{i}}^{*}\psi _{\mathbf{%
i-l_{v}}})
\end{eqnarray}
with $v\in \{x,y\}$ and $\mathbf{l_{x}}=(1,0)$,
$\mathbf{l_{y}}=(0,1)$ and
\begin{equation}
K(\mathbf{n})=\frac{N}{2\pi ^{2}a}\int_{0}^{\pi }dxdy\frac{%
\cos(n_{x}x)\cos(n_{y}y)}{\sqrt{4-2\cos(x)-2\cos(y)}}.
\end{equation}
Here we should point out that Eq.(5) contains 2D integration only\cite
{mertelj}. All dependence on the thickness of the sample appears through the
parameter $\lambda _{eff}=\lambda ^{2}/d$ ($d$ is the thickness of the
sample)\cite{degennes}. This is important difference from the case of
the cylindric sample considered in the Ref.\cite{misko} where the function $K(%
\mathbf{n})$ is essentially different.

\begin{figure}[ht]
\centering\includegraphics[angle=-90,width=4 in]{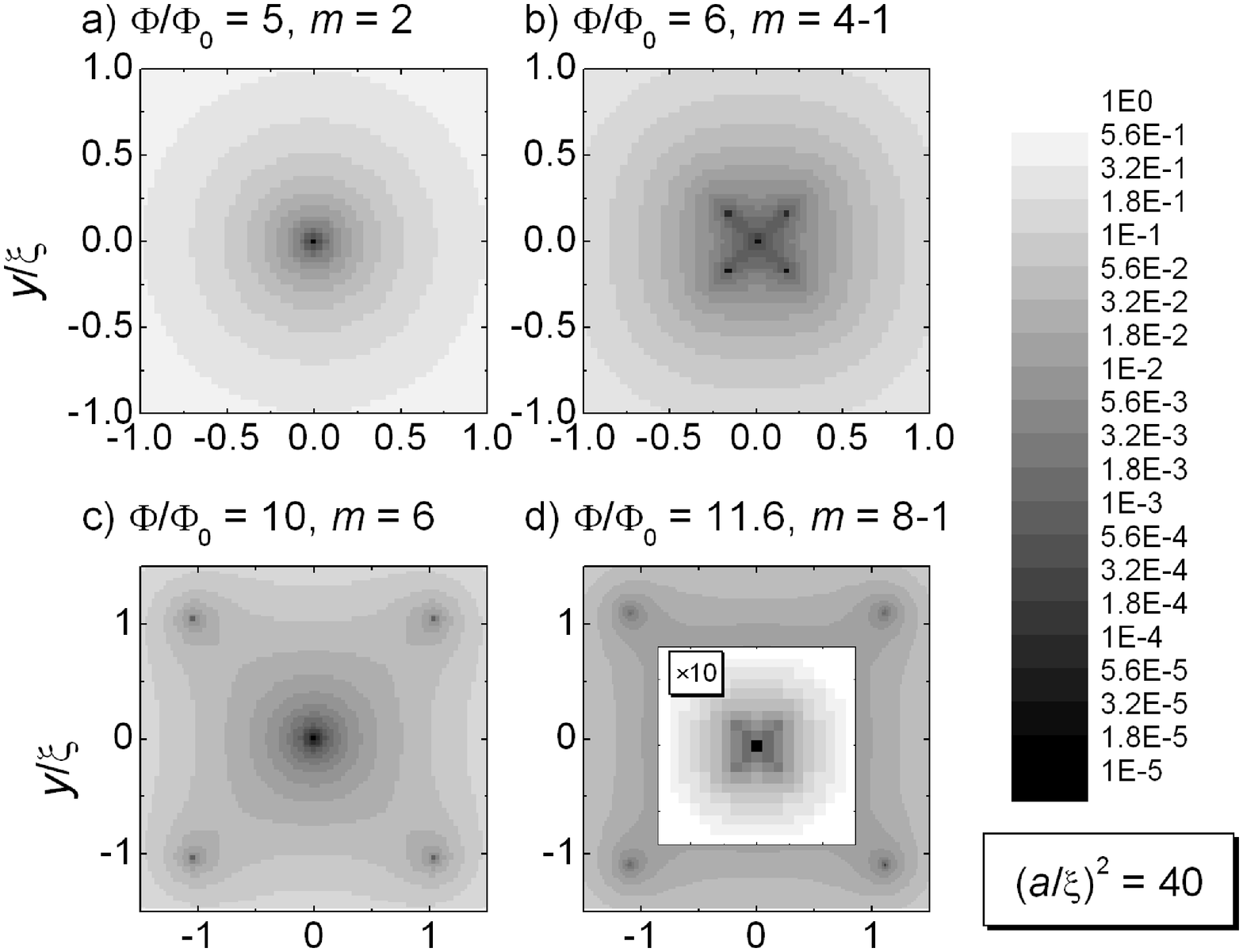}
\caption{The modulus of the order parameter $|\psi |$ at different
magnetic fields. The parameter $\Phi$ is the total external
magnetic flux through the sample. The inset in d) shows the
central region in an expanded scale. Note the logarithmic
intensity scale.}
\end{figure}

The numerical self consistent solution of the problem is obtained by
iterating the solution of the nonlinear equation for the order parameter
Eq.(4) and calculations of the current and the vector potential Eqs.(5,6).
We used two ways of solving Eq.(4). The first is similar to that reported in
ref.\cite{bonca} and corresponds to the iterative solution of the linearized
Eq.(4). The second relies on the fact that Eq.(4) represents the Euler
equation for the free-energy functional with included boundary conditions.
Eq.(4) was therefore solved by the direct minimization of the corresponding
functional using the conjugate-gradient method. Both techniques gave
identical results.

\begin{figure}[hp]
    \centering\includegraphics[angle=270,width=4 in]{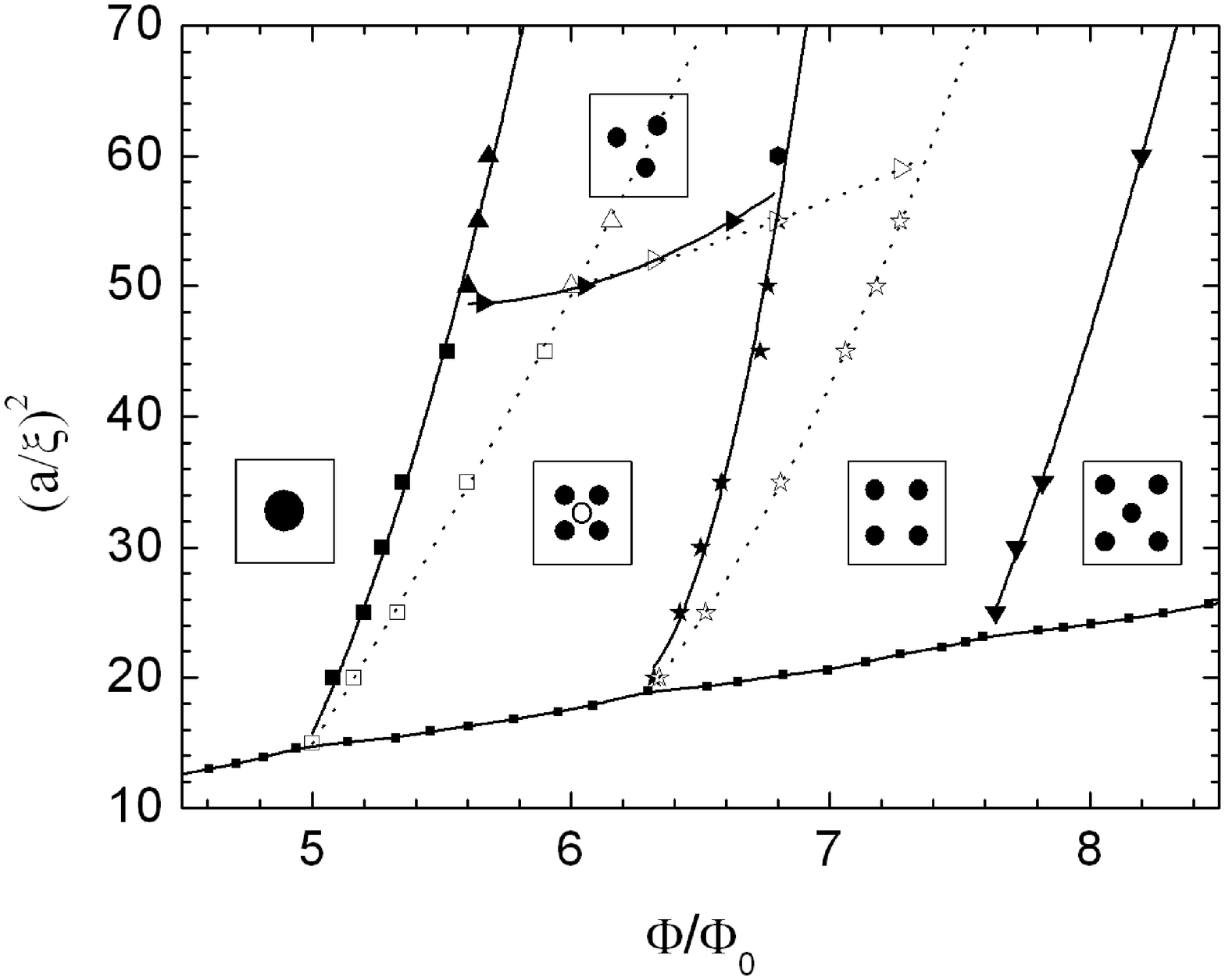}
\caption{The calculated phase diagram. Different phases are marked
with icons schematically indicating the vortex pattern where the
full dot represents a vortex, the open dot reperesents an
antivortex and the larger full dot represents a double vortex. The
full symbols and continous lines represent the phase boundaries
for $\kappa _{eff}=\infty $ while the open symbols and dotted
lines reperesent the phase boundaries for $\kappa _{eff}=1$. In
the later case only the phase boundaries of the region with the
total vorticity 3 are shown.}
\end{figure}

\begin{figure}[hp]
    \centering\includegraphics[angle=270,width=4 in]{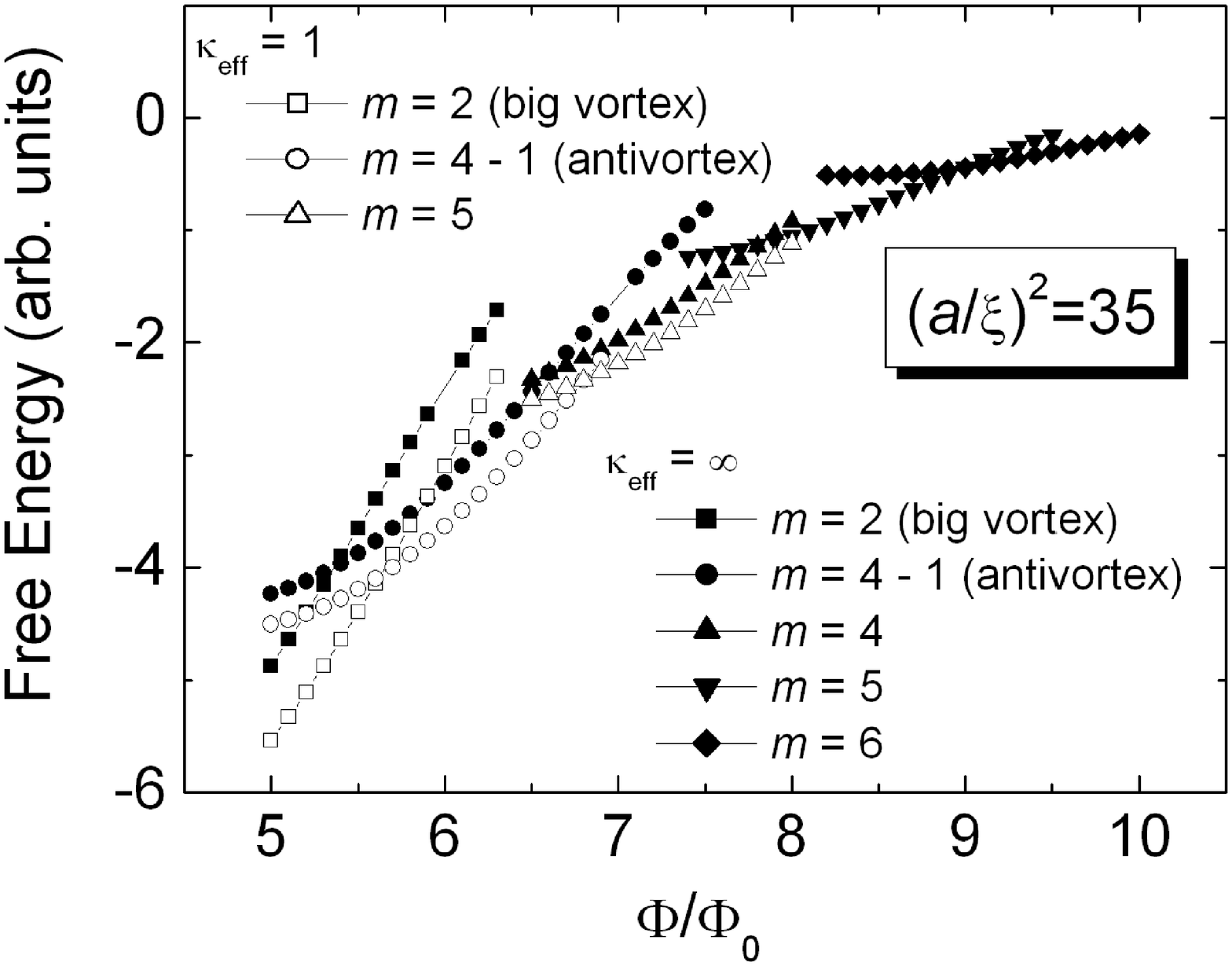}
    \caption{The free energy as a function of magnetic field for
different vorticities $m$.}

\end{figure}

\section{Results}
We investigate first the phase diagram in the regions where the
solution with one antivortex and four vortices and one antivortex
and eight vortices has been reported. In Fig.1 we present the
change of a spatial pattern of the modulus of the order parameter
$|\psi |$ when the vorticity changes from $m=2 $ (Fig.1a) to
$m=4-1$ (Fig.1b) and from $m=6$ (Fig.1c) to $m=8-1$ (Fig.1d).

In the cases $m=2$ and $6$ (Fig.1a and Fig.1b) we observe a giant
double vortex in the centre. In the cases with $m=4l-1$ with
$l=1,2$ instead of the giant triple vortex the symmetry induced
square pattern of four vortices with the antivortex in the middle
forms. The results of the calculations show that the region of the
phase diagram where the symmetry induced antivortex solution has
the lowest energy is broader than expected from the solution of
the linearized GLE. As it is shown in Fig. 2 for $\kappa
_{eff}=\infty $ the antivortex phase is stable up to $(a/\xi
)^{2}\sim 55$, depending on $\Phi /\Phi _{0}$.
For a finite $\kappa _{eff}$ this region shifts to the higher field as $%
(a/\xi )^{2}$ increases (see Fig.2). In the Fig.3 we presented the
calculated value of the sample free energy as a function of external field
for $(a/\xi )^{2}=35$ and for two values of $\kappa _{eff}=1,\infty $.
Topological phase transitions with the changes of the total vorticity $%
\Delta m=1$ are clearly seen. Reduction of $\kappa _{eff}$ leads
to the shift of the transition point to the higher field.

\begin{figure}[ht]
    \centering\includegraphics[angle=270,width=4 in]{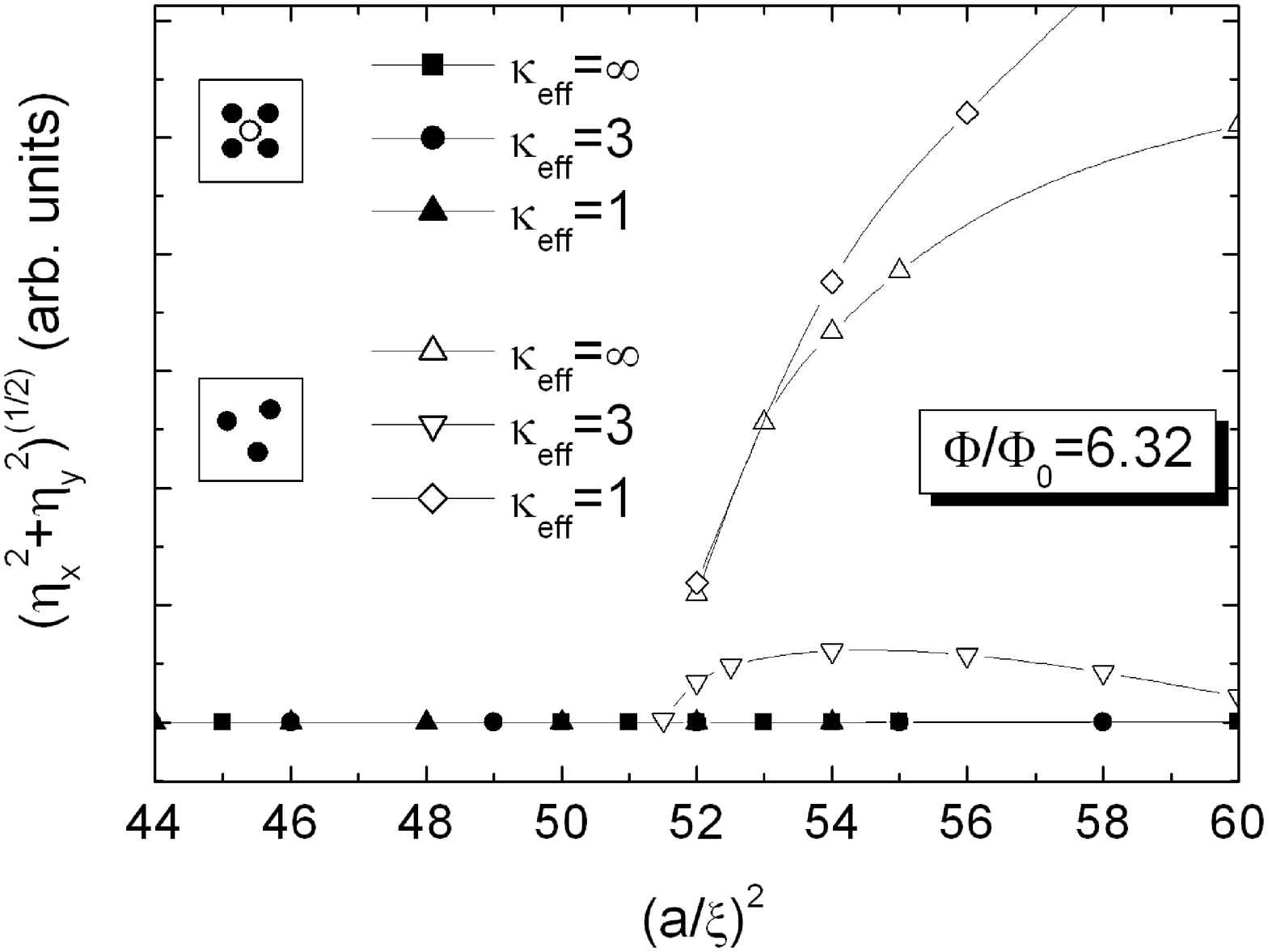}
\caption{The magnitude of the order parameter around the
transition where one vortex anihilates with the antivortex as a
function of $(a/\xi )^{2}$ at the constant magnetic field.}
\end{figure}

The interesting behavior is observed when the external field is fixed and $%
(a/\xi )^{2}$ increases. Close to the $H_{c2}$ the lowest minimum of the
free energy corresponds to the solution with the vorticity $m=4-1$ with the
antivortex in the center of the square. Present calculations do not confirm
the existence of the giant-vortex solution with $m=3$ in this region of the
phase diagram as reported previously \cite{bonca}. The difference is due to
increase of the number of discrete points $N$ enabling detection of the
antivortex. With increase of $(a/\xi )^{2}$ away from the $H_{c2}$ line the
phase transition to the multivortex state with the same vorticity ($m=3$)
and a lower symmetry takes place (see Fig. 2). In general, the free energy
depends on the vorticity $m=n_{+}-n_{-}$ and the total number of vortices in
the system $n=n_{+}+n_{-}$. The transition at $(a/\xi )^{2}\sim 55$ and $%
\Phi /\Phi _{0}\sim 5.5$ takes place at the constant vorticity $m=3$ with
the change of $n$ from 5 to 3. The transition is therefore not only
characterized by an order parameter, but also by the change of the number of
vortices at the constant total vorticity $m$, suggesting that the transition
is close to the first order. This statement is confirmed by the observation
that above the transition point, $(a/\xi )>(a/\xi )_{crit}$, both solutions
with $m=3$ and $m=4-1$ coexists. Since near the transition the free-energy
difference between the phases with the same vorticity $m$ and different $n$
is small it is difficult to determine the phase boundary between phases with
$m=4-1$ and $m=3$ accurately. The transition could be easier observed by
calculating the two component order parameter $\eta _{x}=\int x|\psi
(x,y)|^{2}dxdy$, $\eta _{y}=\int y|\psi (x,y)|^{2}dxdy$ shown in Fig.4.%

\begin{figure}[ht]
    \centering\includegraphics[width=2.8 in]{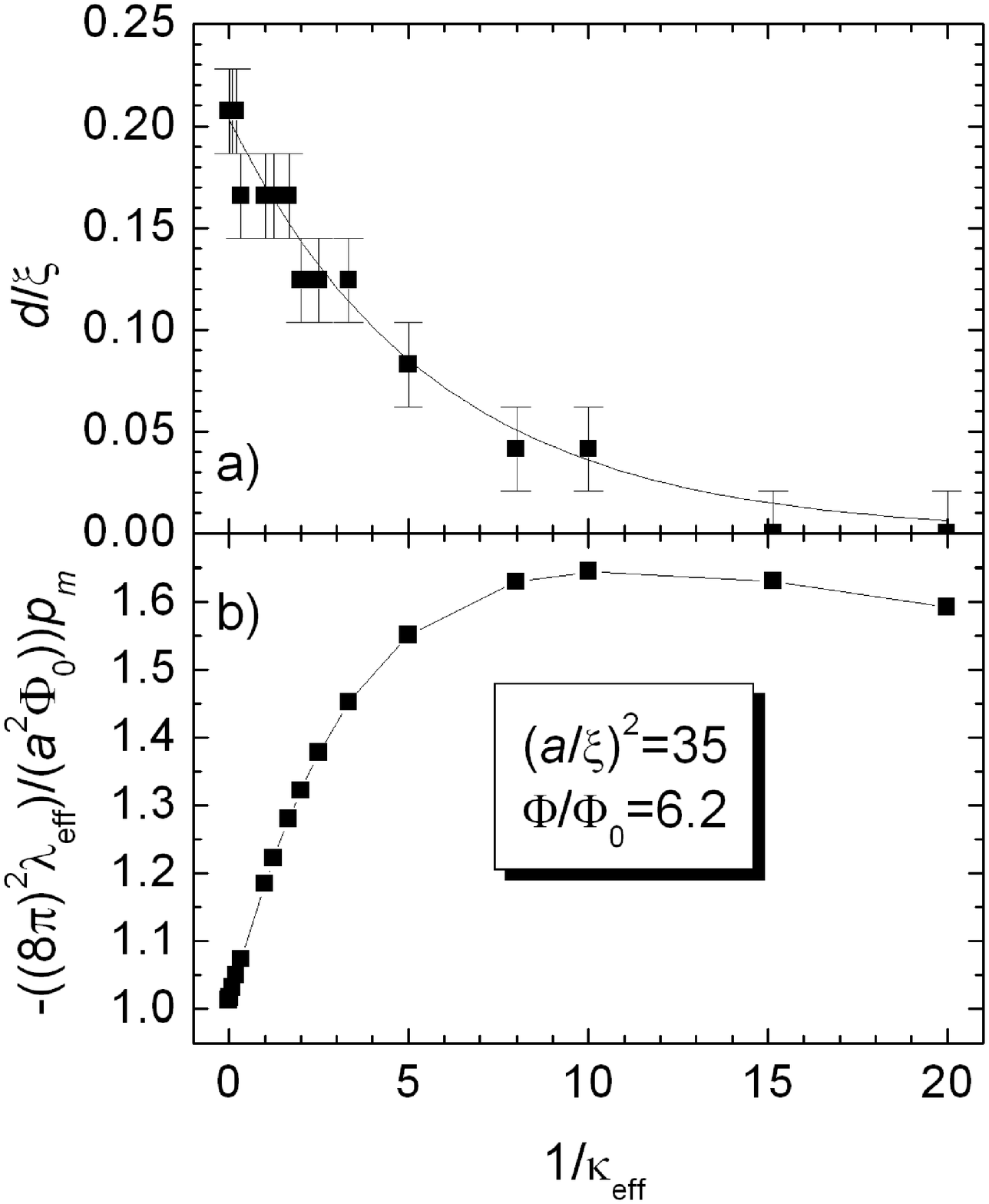}

    \caption{The vortex-antivortex distance
    a) and the magnetic moment of the sample b) as functions of the parameter $%
    1/\kappa _{eff}$. In (a) error bars represent the grid spacing and
    the solid line the exponential fit discussed in the text.}
\end{figure}

Close to the $H_{c2}$ line the repulsion of vortices from the
boundaries and
attraction of the 4 vortices to the antivortex stabilizes the phase with $%
m=4-1$ and small vortex-vortex distances. At smaller value of $\xi $ the
repulsion from the boundaries decreases and one vortex annihilates with the
antivortex. As a result, the repulsion between the remaining vortices
increases leading to an increase of the order parameter with a further
decrease of $\xi$.

Increasing the external field up to $\Phi /\Phi _{0}=11.6$ leads
to the stabilization of the phase with total vorticity $m=7$. Near
the $H_{c2}$ line similar to
the case with $m=4-1$ the solution with $m=8-1$ is realized (Fig.1d). When $%
(a/\xi)^{2}$ increases in a complete analogy to the case with $m=4-1$ the
phase transition to the phase with $m=7$ and a similar order parameter takes
place. Here also, both solutions with $m=8-1$ and $m=7$ coexists above $%
(a/\xi)_{crit}$ indicating that the transition is close to the first order.
We believe that the situation is quite general for the case of arbitrary $%
m=4l-1$ for $l=1,2,3,...$.

At the end we would like to discuss the dependence of the
stability of the antivortex phase at small $\kappa_{eff} $.
According to the arguments of Ref.\cite {misko}, at small $\kappa
$ the vortex-vortex interaction changes the sign making the
antivortex phase more stable. As a result, the average distance
between vortices in the middle of the square increases as well. In
order to verify this conjecture for the thin film sample we plot
in Fig.5 the vortex-antivortex distance $r_{0}$ as a function of
$1/\kappa _{eff}$. The distance decreases with the decreasing
$\kappa _{eff}$. For $\kappa _{eff}<0.1$ the distance is smaller
than the grid spacing $a/N$ so we can not resolve separate
vortices. We find that $r_{0}\propto \exp {(-\Lambda /\lambda
_{eff})}$ {with }${\Lambda \sim a}$. The situation is just
opposite to that reported in ref.\cite{misko}. We believe that in
the case of the thin film of the square shape the reduction of
$\kappa $ does not stabilize the phase with the antivortex.

\begin{figure}[ht]
    \centering
        \includegraphics[width=2.8 in]{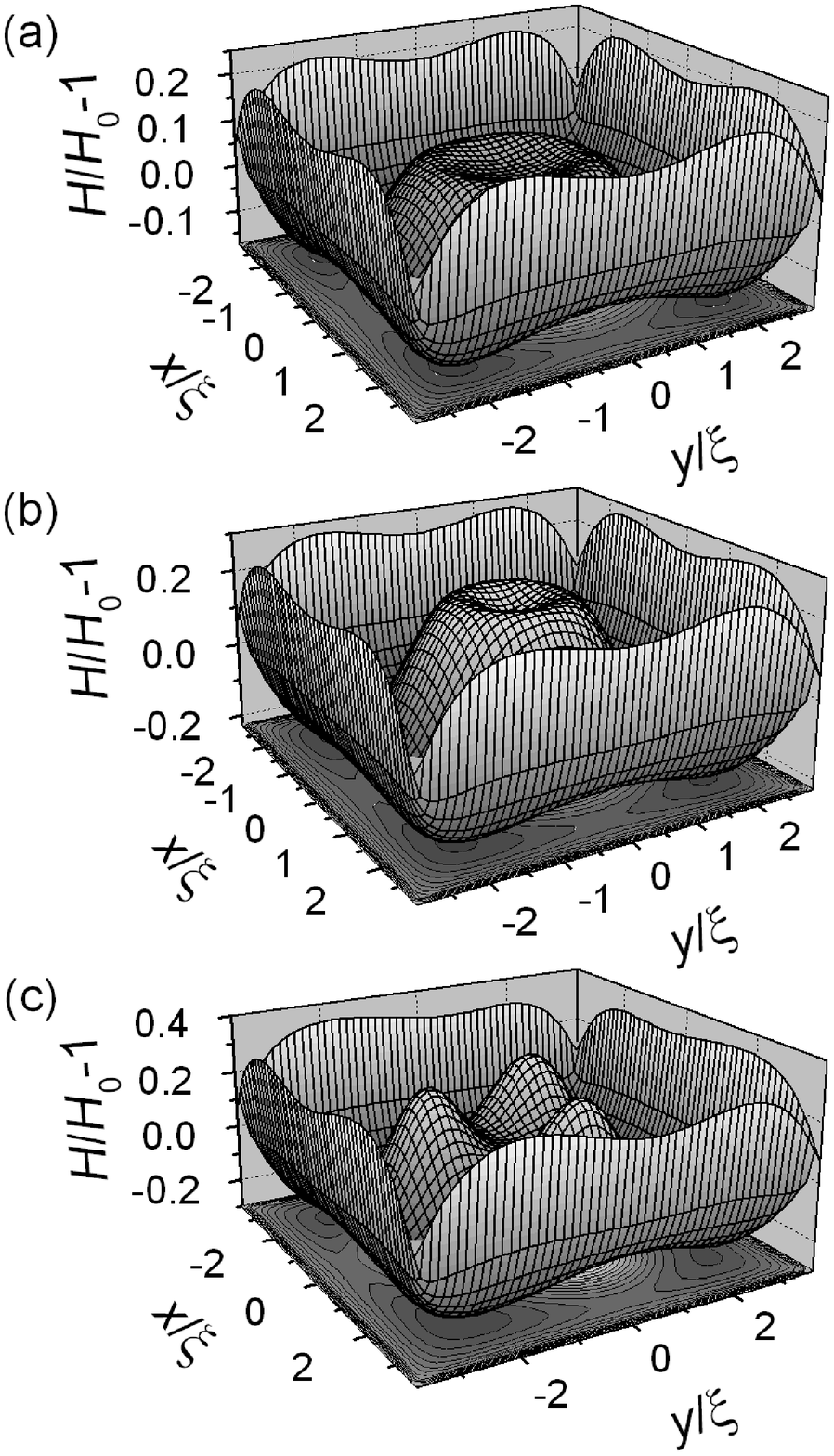}
    \caption{The magnetic field in the film in the case
    of (a) giant vortex with $m=2$, (b) the antivortex solution with
    $m=4-1$ and (c) three separate vortices with $m=3$. Here $H_{0}$
    is the external magnetic field.}
\end{figure}

It is interesting to note differences between samples of different
shapes. For the cylindric shape the giant vortex phase with any
vorticity is always stable close to the $H_{c2}$
line\cite{Schw,Peet}. According to Ref.\cite {misko} for the
mesoscopic triangle the giant vortex state with $m=2$ is
metastable and the solution with the antivortex ($m=3-1)$ is
stable. For the case of the square shape the giant vortex solution
with $m=3$ is \emph{never} stable for $\kappa _{eff}\ge 0.1$. For
$\kappa _{eff}<0.1$ the limited grid spatial resolution prevented
us to distinguish the solution with the antivortex from the
possibly (meta)stable giant-vortex solution.

Finally, let us discuss the possibility to detect the state with the
antivortex experimentally. Calculation of the magnetic field in the sample
shows that the magnetic field has a local minimum in the center of the
sample also for the giant vortex solution with $m=2$. The local minimum
observed for the antivortex state with $m=4-1$ is therefore not due to the
antivortex formation (Fig.6) but due to a particular distribution of the
current in the sample. Therefore, imaging of the magnetic field distribution
cannot provide an evidence for the antivortex. The magnetic field for the
multivortex solution with $m=3$ has 3 well separated maxima that break the
four fold rotational symmetry of the sample allowing a direct imaging of
vortices. Since the antivortex state cannot be detected directly the
observation of a hysteresis in the vicinity of the transition line from the $%
m=4-1$ antivortex state to the $m=3$ multivortex state could suggest that
the symmetric phase is indeed the phase with the antivortex.

\begin{acknowledgments}
The authors wish to thank J. Bonca,  A.S. Alexandrov and V.V. Moshchalkov for
useful correspondence and discussions.
\end{acknowledgments}

\end{document}